\documentclass[12pt]{article}
\usepackage[dvips]{graphicx}

\newcommand{\GeV}{\ensuremath{\mathrm{Ge\kern -0.12em V}}}
\newcommand{\TeV}{\ensuremath{\mathrm{Te\kern -0.12em V}}}

\begin{document}
\begin{center}

{\Large\bf Photon-induced Background for Dilepton Searches\\
and Measurements in pp Collisions at 13 TeV}
\vspace{1cm}

{\sc 
Dimitri Bourilkov
}

{\small bourilkov@mailaps.org
}

{\sl
Physics Department, University of Florida, P.O. Box 118440\\
Gainesville, FL 32611, USA}

\end{center}

\begin{abstract}

The production of high invariant mass opposite sign lepton pairs in
proton-proton collisions at the LHC is dominated by the Drell-Yan
process. In addition to this photon or Z exchange mediated mechanism,
gamma-gamma collisions, where photons radiated by the incoming protons
collide, can produce lepton pairs. This is an important additional
source of background for high mass resonant (like Z') or non-resonant
(like contact interactions) searches. In this paper detailed
calculations of the Drell-Yan and photon-induced cross sections in the
typical acceptance of a multi-purpose LHC detector at center of mass
energy 13~TeV are presented.
The hint for a diphoton excess at a mass around 750~GeV, reported by
the ATLAS and CMS experiments from the analysis of the 2015 data at 13
TeV, raises the possibility that such an excess could be produced
through gamma-gamma collisions. A good theoretical understanding and
measurements in the dilepton channels at these energies can help to
elucidate this production hypothesis.
\end{abstract}

\section{Introduction}

The Standard Model (SM) of particle physics has been tested
extensively at collider energies. Searches for new physics phenomena
in dilepton (electron or muon) final states provide clean signatures.
The CMS and ATLAS collaborations at the LHC have searched for
resonant or non-resonant effects beyond the SM, see
e.g~\cite{Chatrchyan:2011wq,Khachatryan:2014fba,Aad:2014cka,Aad:2014wca}.

The backgrounds at high invariant masses are dominated by the
Drell-Yan (DY) process of opposite sign lepton pair production,
mediated through photon or Z exchange from the initial partons in the
incoming protons. An alternative route to produce lepton pairs is in
gamma-gamma collisions, where photons radiated by the incoming protons
collide. To calculate this process, usually labeled photon-induced (PI)
background in various searches, we need parton density functions
(PDF) of the protons including the photon component. A similar process,
two-photon collisions, where the photons originate from electrons and
positrons, has been studied extensively e.g. at LEP2~\cite{LEP2yb}.

The hint for a diphoton excess~\cite{ATLASgg,CMS:2015dxe}
at a mass around 750~GeV, reported by the ATLAS and CMS collaborations
from the analysis of the 2015 data at 13 TeV, has sparked a flurry of
phenomenological activity. One hypothesis is that a spin 0 scalar is
produced through gamma-gamma collisions~\cite{Csaki:2016raa}. Clearly
a good theoretical understanding and measurements at these energies
can help to elucidate this hypothesis.

\section{The Photon PDF}

Quantum Electrodynamics (QED) introduces corrections to the parton
evolution. As a result, we have to include photon parton distributions
$\gamma(x,Q^2)$ for the proton and the neutron, and part of the
proton (neutron) momentum is carried by the photons. The PDF
depends on the parton momentum fraction - Bjorken $x$, and the
momentum transfer $Q^2$. 

In~\cite{Martin:2004dh} a global analysis of deep inelastic and hard
scattering data was performed and the photon PDF was extracted. It
is available from the LHAPDF
library~\cite{Bourilkov:2003kk,Whalley:2005nh,Bourilkov:2006cj}
as MRST2004qed.
This PDF set uses a model for the starting distribution at momentum
transfer $Q_0^2\ =$~1~GeV$^2$. For the photon radiation cutoff there
are two options:
\begin{enumerate}
 \item Cutoff at the current quark masses of 6 (10) MeV for up (down)
       quarks. We will label this PDF as Mem0 (corresponding to
       member 0 in the LHAPDF library).
 \item Cutoff at the constituent quark masses of 300 MeV for up and down
       quarks. We will label this PDF as Mem1 (corresponding to
       member 1 in the LHAPDF library). This choice reduces the momentum
       carried by the photon, as higher masses produce less radiation.
\end{enumerate}
The authors of~\cite{Martin:2004dh} argue that, photon radiation being
a perturbative QED effect, current quark masses are more appropriate;
correspondingly this option is the default member~0.
They support this argument with an analysis of the then available ZEUS
collaboration data~\cite{Chekanov:2004wr} for the reaction: 
\begin{equation}
       ep \rightarrow e\gamma X
\end{equation}
measured in a narrow range around $x_{\gamma} \sim 0.005$.
This is the most direct measurement of the photon PDF, as the photon
contribution is at the leading perturbative order.

In~\cite{Ball:2013hta} a different starting photon parametrization
and a combined quantum chromodynamics QCD and QED evolution is used.
In addition, data on Drell-Yan, W and Z production from the LHC is
used to constrain the photon PDF. The LHC data places relatively
weak constraints, resulting in large PDF uncertainties. We will
use NNPDF23qed, as provided by LHAPDF.

In~\cite{Schmidt:2015zda} a generalization of the MRST approach is
applied to derive CT14QED sets. The initial photon distribution is
defined by the initial photon momentum fraction $p_0^{\gamma}$ at scale
$Q_0\ =\ $1.295~GeV. Newer ZEUS data on deep inelastic scattering with
isolated photons~\cite{Chekanov:2009dq} is used to constrain the
photon PDF. The authors find
\begin{equation}
       p_0^{\gamma} \le 0.14\%
\end{equation}
at 90\% confidence level (CL). They conclude that the Mem0 option of
MRST2004qed, using current quark masses, is ruled out by the data,
and recommend using initial $p_0^{\gamma}$ values between 0 and 0.14\%.
The constituent quark masses option of MRST2004qed is allowed, as well
as the initial NNPDF parametrization at low $Q^2$. At high $Q^2$ the
CT14QED sets are compatible with MRST2004qed (Mem1), while the
NNPDF23qed set shows sizeable deviations both at low and high x
values. The implications of this for LHC predictions will be analyzed
in detail in the next section. 

In Drell-Yan, W and Z production at the LHC the photon contribution
is suppressed by a factor O($\alpha / \alpha_s)$ compared to the
canonical quark-antiquark contribution, thus providing relatively
weak constraints on the photon PDF, especially where high precision
data is scarce, e.g. at high masses or high x values. The sensitivity
can be improved in measurements where the photon contribution is
enhanced by selecting exclusive dimuon pair production in elastic,
single dissociative and double dissociative pp
collisions~\cite{Chatrchyan:2011ci,Khachatryan:2016mud}. These CMS
measurements are used in~\cite{Ababekri:2016kkj} to update the CT14QED
analysis. The authors conclude that
\begin{eqnarray}
p_0^{\gamma} \le 0.09\%\ at\ 68\%\ CL \\
p_0^{\gamma} \le 0.13\%\ at\ 90\%\ CL
\end{eqnarray}
consistent with the ZEUS data analysis. At 250~GeV a CT14QED PDF with
momentum fraction 0.09\% is compatible with MRST2004qed Mem1 and
NNPDF23qed, while MRST Mem0 is higher. Above 1~TeV NNPDF23qed exceeds
all other PDFs (even Mem0), while Mem1 is compatible with CT14QED
throughout.

\begin{figure}[ht]
\centerline{\resizebox{0.95\textwidth}{14.0cm}{\includegraphics{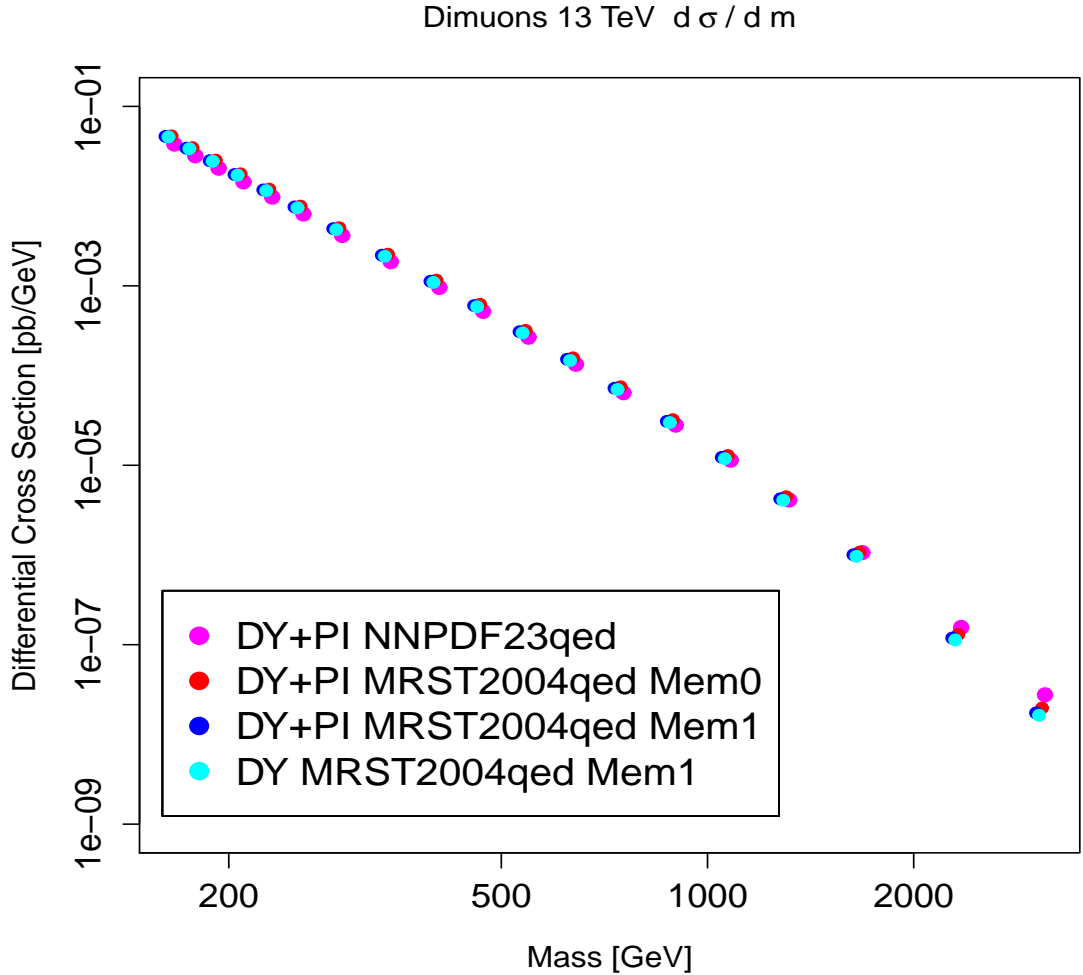}}}
\vspace*{-3pt}
\caption{Differential cross sections for DY and PI at 13~TeV in the
         search acceptance.}
\label{fig:fig01}
\end{figure}

\section{Results}

The calculations for the Drell-Yan process and the photon-induced
background are performed with the program {\tt FEWZ}~\cite{Li:2012wna}.
The $G_{\mu}$ input scheme is used.
The PDFs considered in this study are MRST2004qed and NNPDF23qed.
Full electroweak corrections at next to leading order (NLO) are
computed (the flag {\tt EW control = 0} is used).
QCD effects are computed at leading order (LO).
When the PI background is included we label the results \mbox{DY+PI}.
When only the DY process is included we label the results \mbox{DY}.
Most of the presented plots show ratios of cross sections where
some uncertainties cancel, and the influence of missing higher orders
is reduced.
Calculations for dimuons in the acceptance of a generic general
purpose LHC experiment are presented: both outgoing leptons are
required to have pseudorapidity $|\eta| < 2.4$.
The results for dielectrons are very similar.

We distinguish two cases:
\begin{enumerate}
 \item ``Search'' acceptance: relatively hard cuts suitable for high
       masses are used - cuts on the transverse momenta $p_T > 50$~GeV
       for both leptons are applied.
 \item ``DY'' acceptance: relatively soft cuts suitable also for low
       masses are used - asymmetric cuts on the transverse momenta
       $p_T > 20\ (10)$~GeV for the harder (softer) lepton are
       applied. These cuts allow precision measurements both above
       and below the Z peak.
\end{enumerate}

In Figure~\ref{fig:fig01} the differential cross sections as function
of mass are shown. The substantial differences at higher masses are
visible even in this double logarithmic scale plot.  As the validity
range of MRST2004qed is limited by $Q_{Max}$~=~3162.28~GeV, we show
plots up to this value. This covers early searches but will not be
sufficient with the accumulation of LHC luminosity at 13~TeV. It is
possible to extrapolate to higher masses, or alternatively use the
newest photon PDFs like CT14QED.

It is worth noting that the absolute values of the cross sections
do not agree with the values presented in~\cite{Ball:2013hta},
see e.g. their Figure~23. It appears that the differential cross
section there is not properly normalized, and it exhibits an
unphysical ``kink'' around 1.2~TeV. 

\begin{figure}[htb]
\centerline{\resizebox{0.95\textwidth}{10.5cm}{\includegraphics{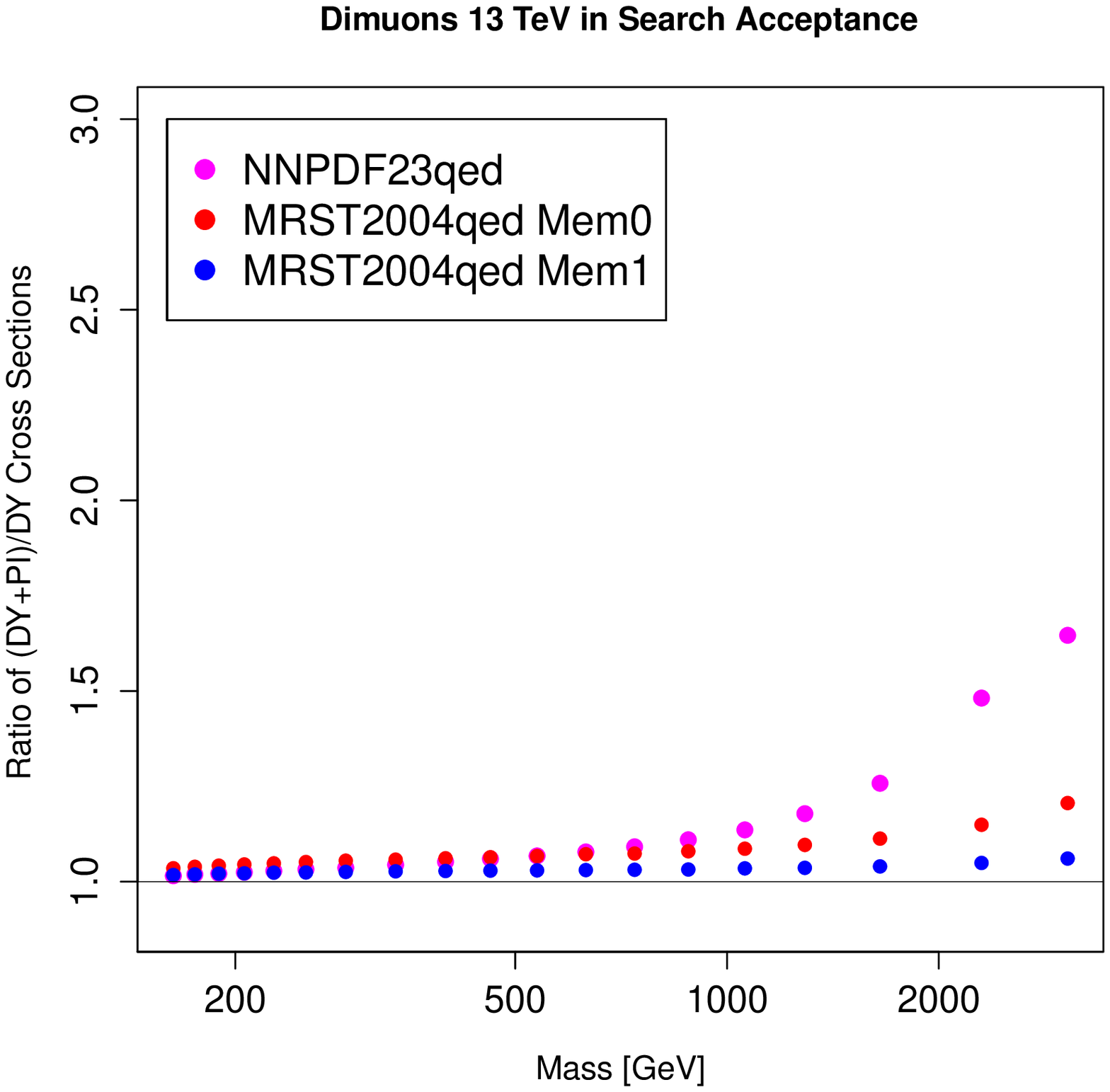}}}
\centerline{\resizebox{0.95\textwidth}{10.5cm}{\includegraphics{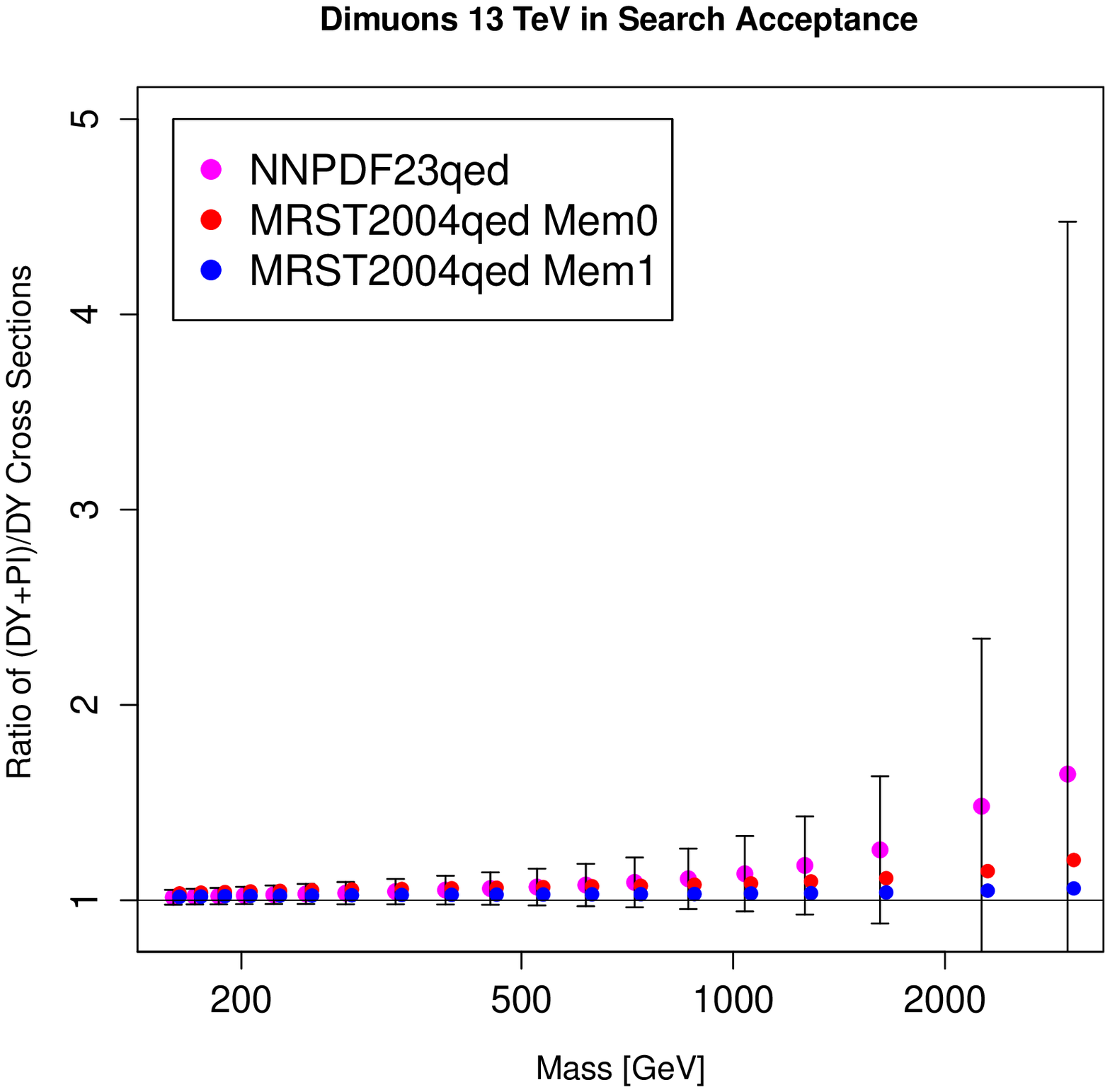}}}
\vspace*{-3pt}
\caption{Top: ratios of cross sections (DY+PI)/DY for different PDFs
         in the Search acceptance.
         Bottom: same, but including the PDF uncertainties for
         NNPDF23qed.}
\label{fig:fig02}
\end{figure}

In Figure~\ref{fig:fig02} the predictions for the cross section ratio
(DY+PI)/DY for different photon PDFs are shown in the Search
acceptance. The bottom plot shows the same ratios including the PDF
uncertainty for NNPDF23qed. As can be seen, the latter predictions
differ substantially from the MRST2004qed results, especially at
masses above 1~TeV. Moreover, the PDF uncertainty exceeds the size
of the PI effect, limiting the predictive power of this calculation.
Clearly this is related to the lack of experimental data at high
momentum transfers, the accumulation of much more data in
Run~2 certainly will help.

\begin{figure}[ht]
\centerline{\resizebox{0.95\textwidth}{10.5cm}{\includegraphics{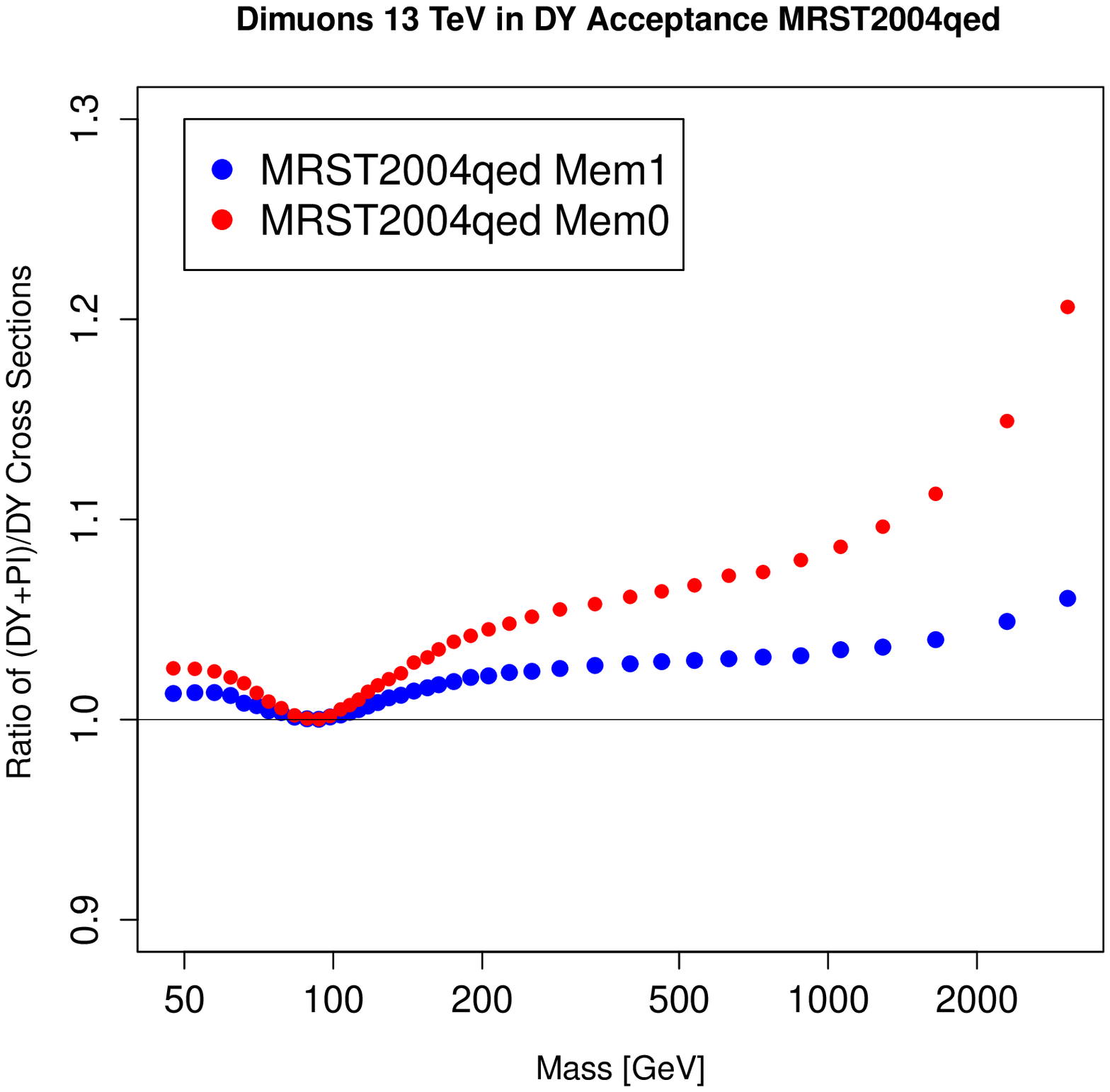}}}
\centerline{\resizebox{0.95\textwidth}{10.5cm}{\includegraphics{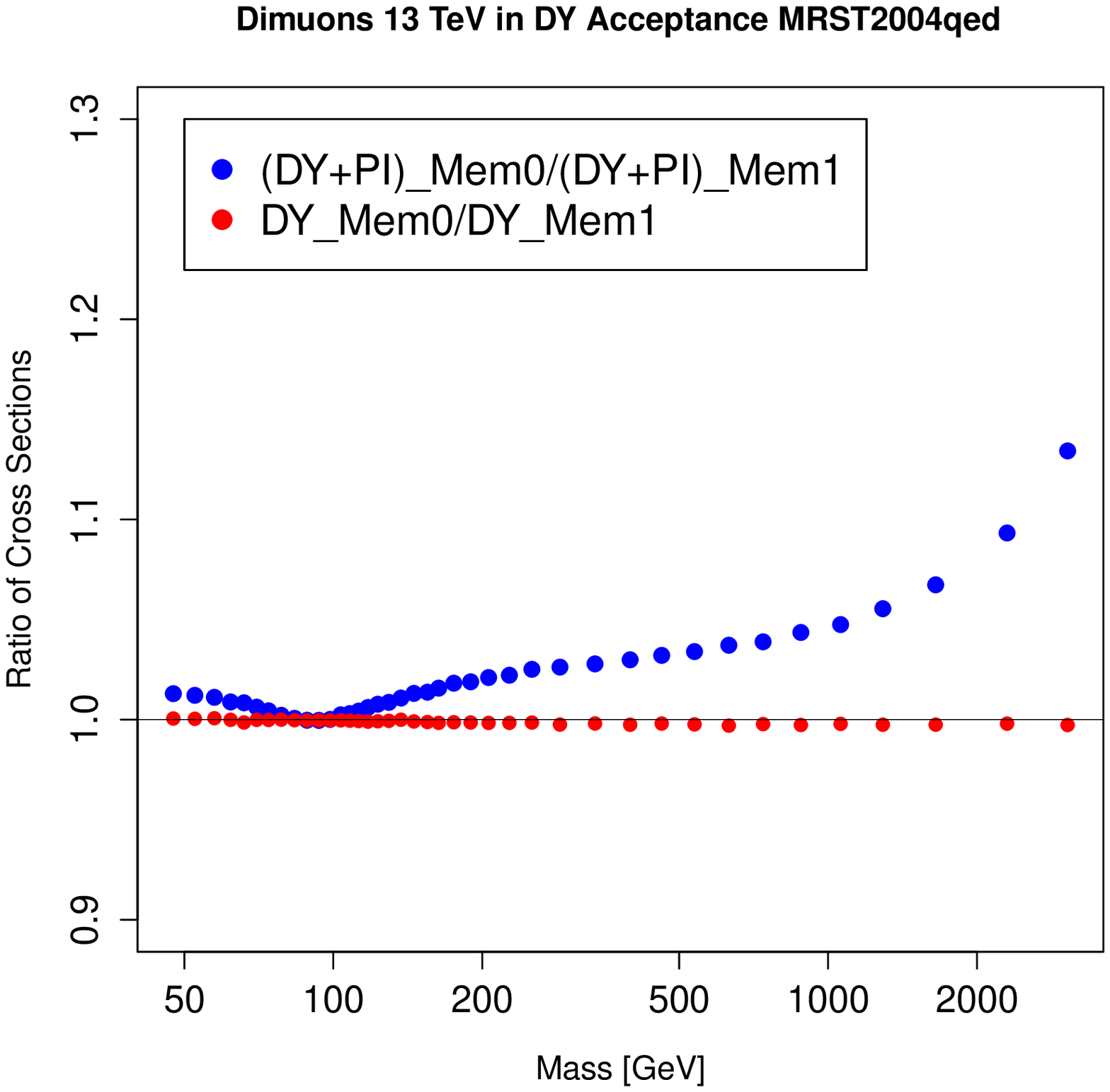}}}
\vspace*{-3pt}
\caption{Top: ratios of cross sections (DY+PI)/DY for the two members
         of MRST2004qed in the DY acceptance.
         Bottom: ratio of (DY+PI) for the two members (blue), and
         ratio of the DY cross sections for the two members (red),
         showing that the DY predictions are almost identical while the
         (DY+PI) predictions differ substantially.}
\label{fig:fig03}
\end{figure}

The cross sections in the DY acceptance for the two members of
MRST2004qed are compared in detail in Figure~\ref{fig:fig03}. The DY
predictions are almost identical, as expected, while the (DY+PI)
results differ substantially. As discussed in the previous section,
the experimental data favor the Mem1 option, producing the smallest PI
effects.

\begin{figure}[ht]
\centerline{\resizebox{0.95\textwidth}{10.5cm}{\includegraphics{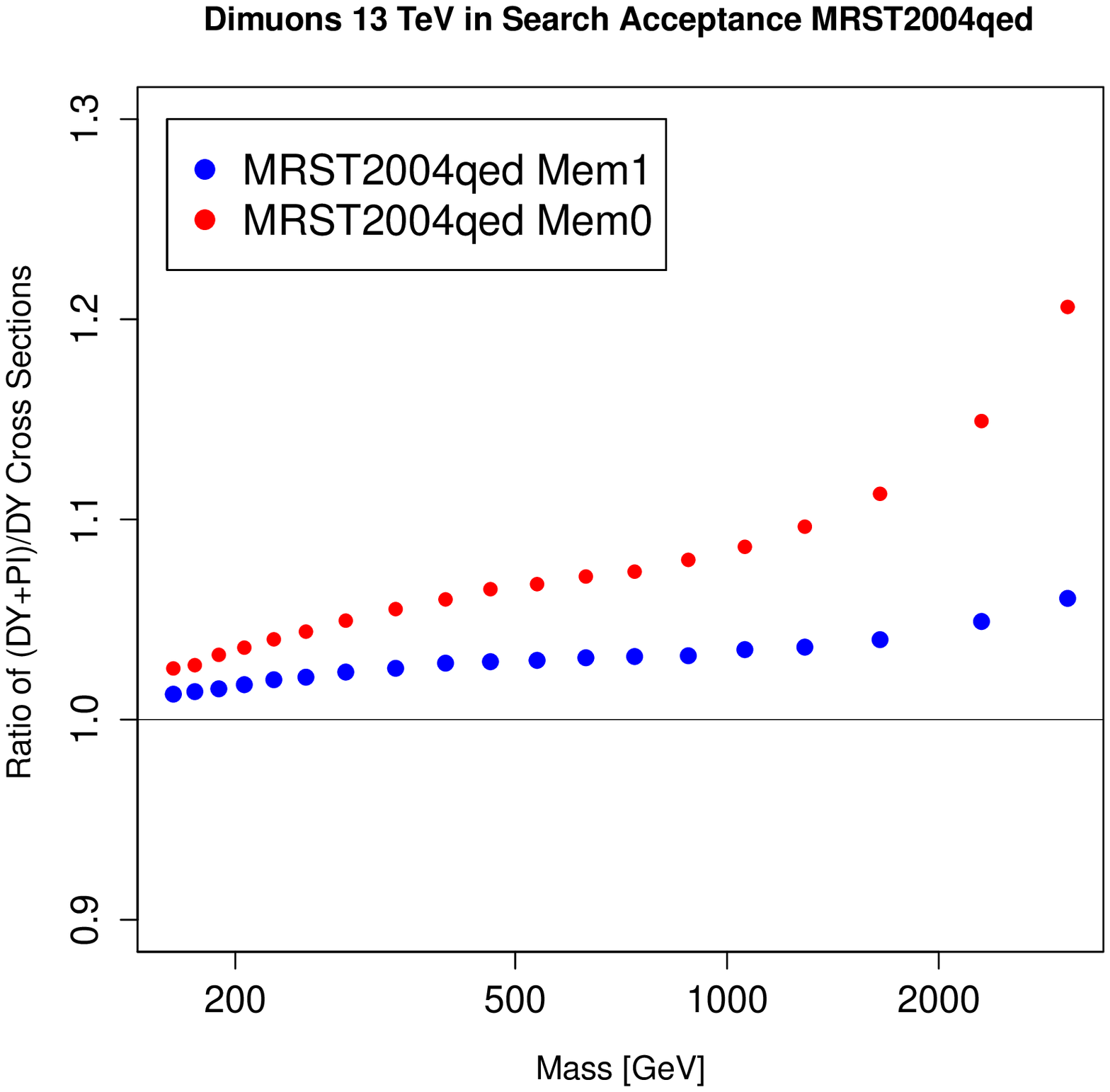}}}
\centerline{\resizebox{0.95\textwidth}{10.5cm}{\includegraphics{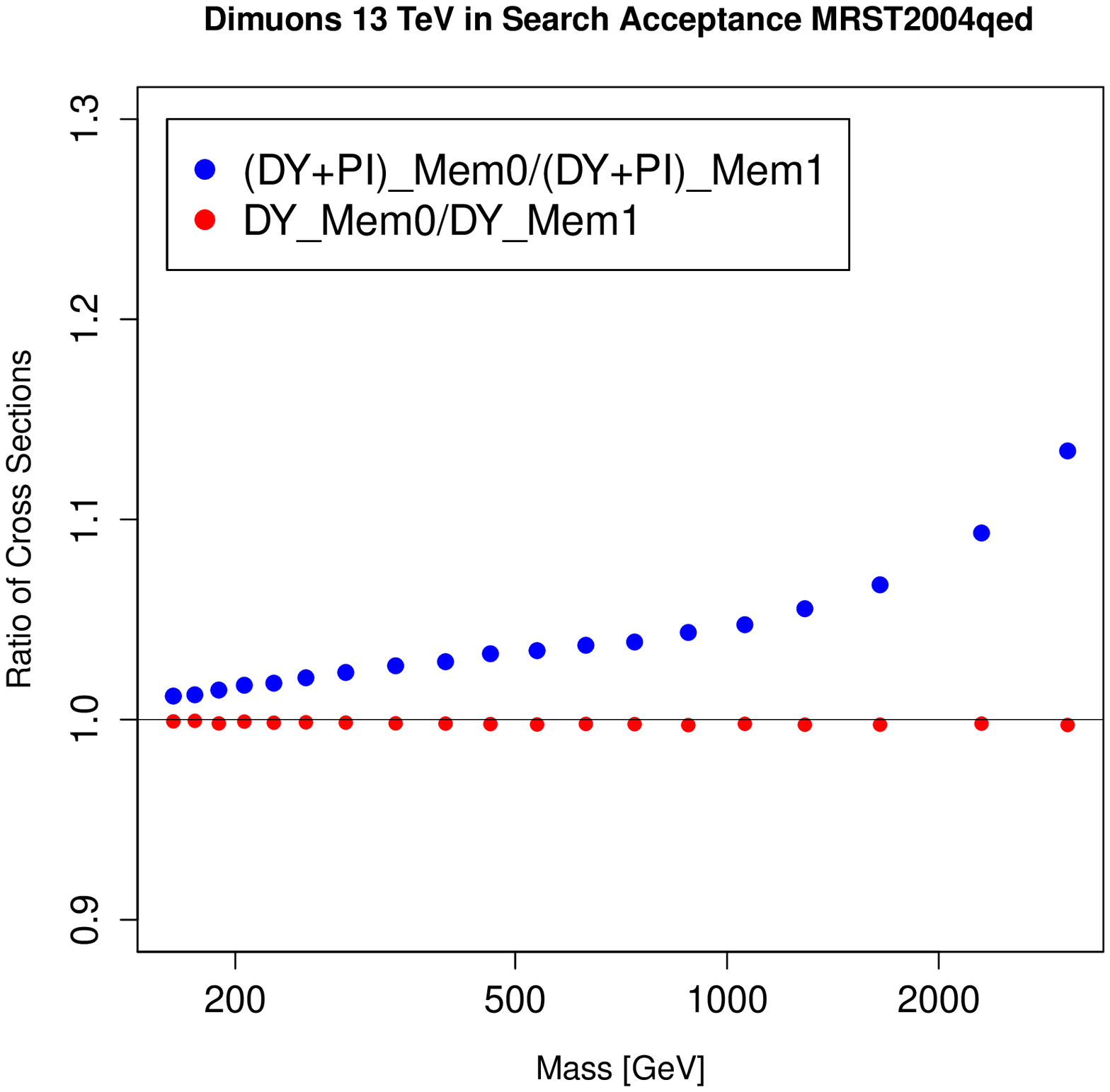}}}
\vspace*{-3pt}
\caption{Top: ratios of cross sections (DY+PI)/DY for the two members
         of MRST2004qed in the Search acceptance.
         Bottom: ratio of (DY+PI) for the two members (blue), and
         ratio of the DY cross sections for the two members (red),
         showing that the DY predictions are almost identical while the
         (DY+PI) predictions differ substantially.}
\label{fig:fig04}
\end{figure}

In Figure~\ref{fig:fig04} the cross sections in the Search acceptance
for the two members of MRST2004qed are compared in detail. Again the
DY predictions are almost identical, as expected, while the (DY+PI)
results differ substantially. As discussed in the previous section,
the experimental data favor the Mem1 option, producing the smallest PI
effects, important for resonant and especially for non-resonant
searches in the TeV region.

\clearpage

\clearpage

\clearpage

\begin{figure}[ht]
\centerline{\resizebox{0.95\textwidth}{10.5cm}{\includegraphics{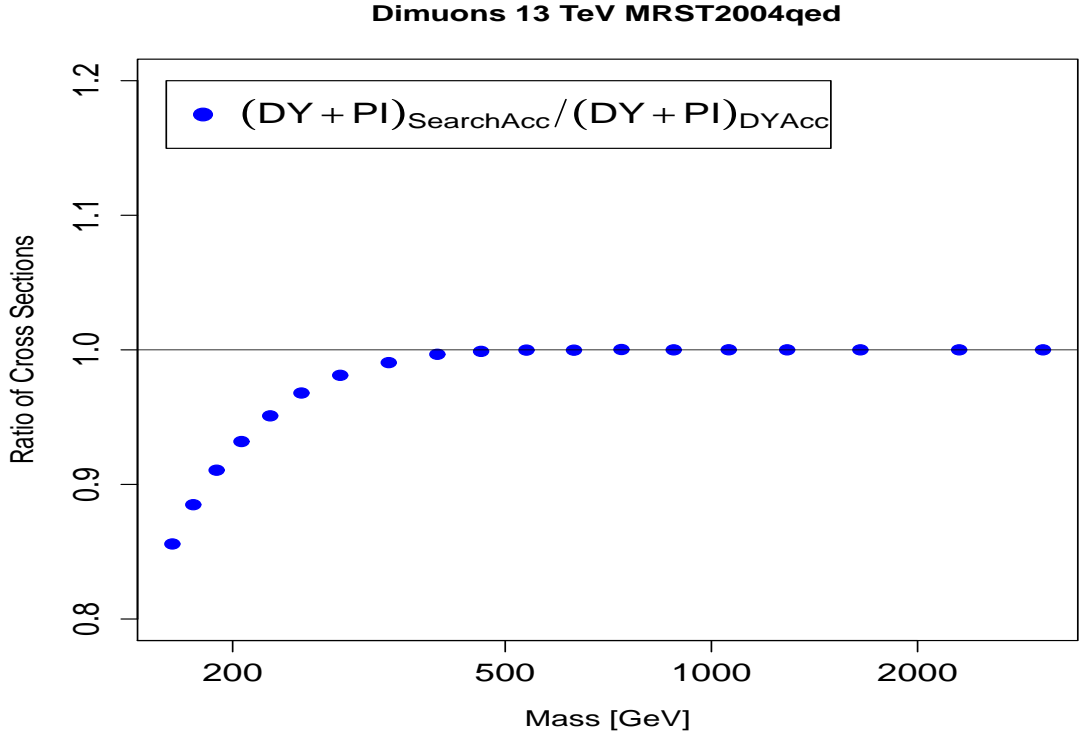}}}
\vspace*{-3pt}
\caption{Ratios of cross sections (DY+PI) for MRST2004qed in the Search
         and DY acceptances. At high mass the ratio approaches unity.}
\label{fig:fig05}
\end{figure}

An important question arises: can the PI background be suppressed by a
set of cuts which does not reduce substantially the signal
efficiency. As shown in Figure~\ref{fig:fig05}, the harder $p_T$ cuts
on the outgoing leptons do not reduce the (DY+PI) cross sections
above $\sim$~400~GeV, so they have no impact on the interesting high
mass region.

\begin{figure}[ht]
\centerline{\resizebox{0.95\textwidth}{10.5cm}{\includegraphics{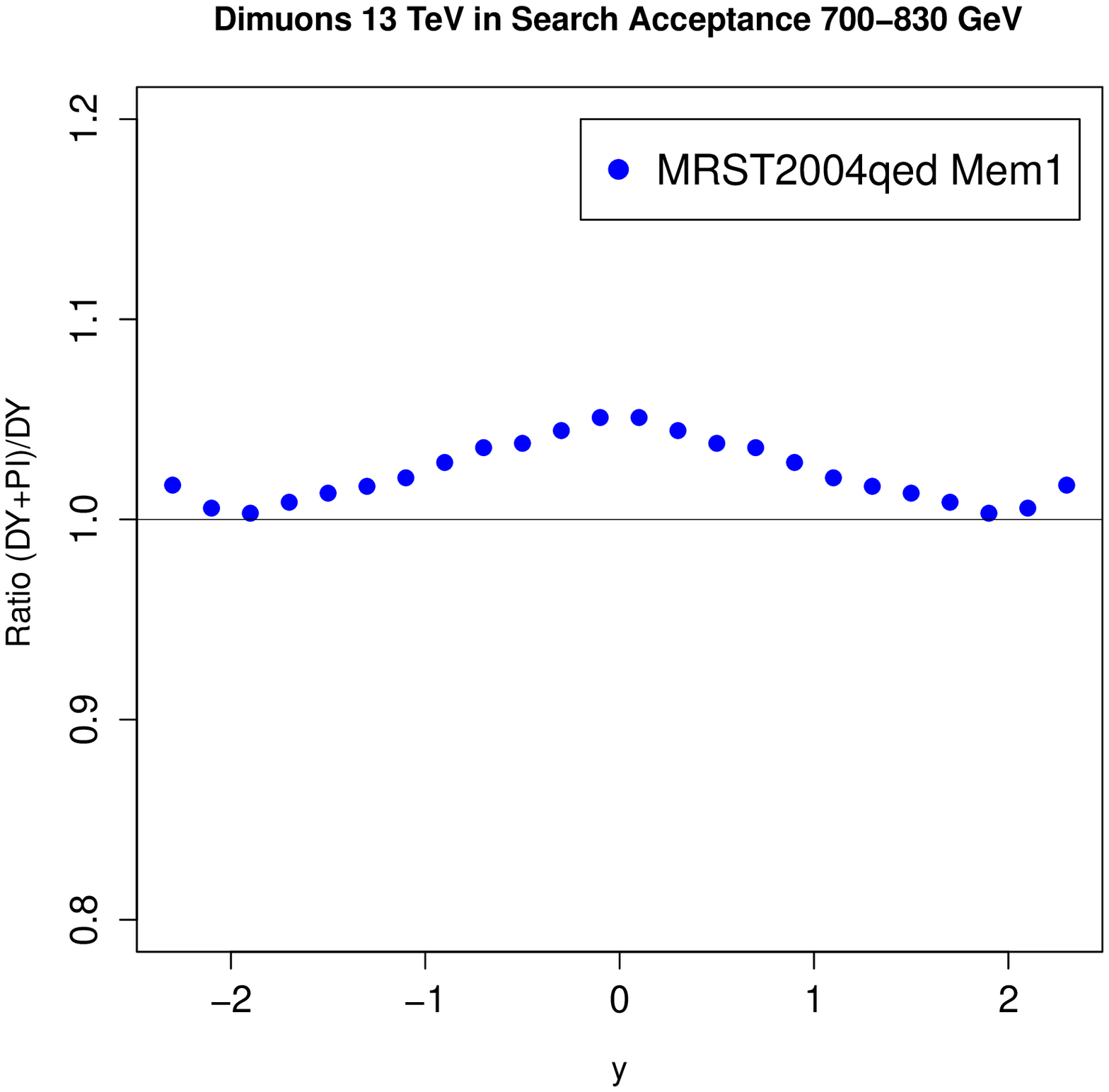}}}
\centerline{\resizebox{0.95\textwidth}{10.5cm}{\includegraphics{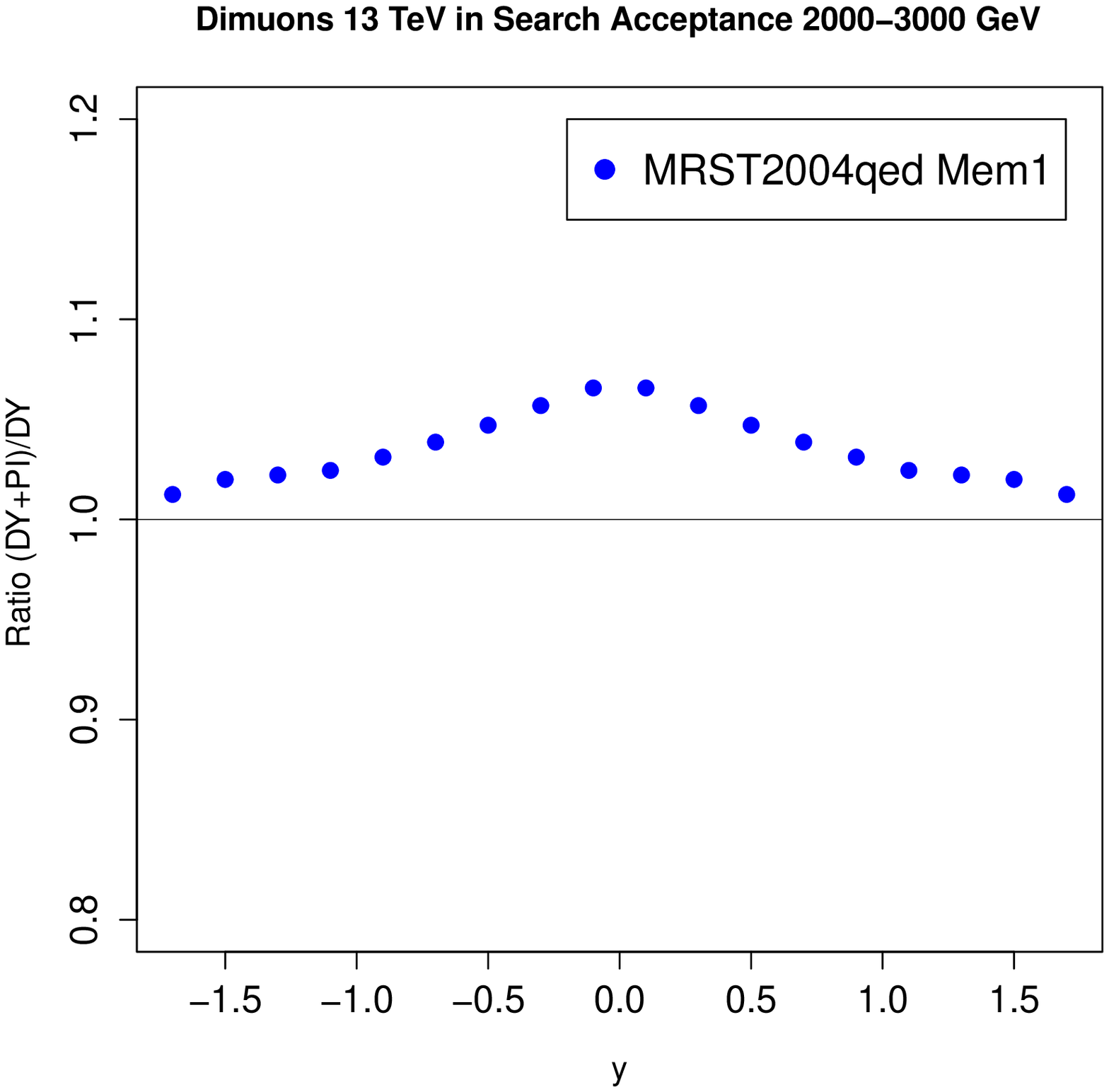}}}
\vspace*{-3pt}
\caption{Top: rapidity distribution for (DY+PI)/DY using MRST2004qed
         Mem1 for invariant masses 700--830~GeV.
         Bottom: rapidity distribution for the same ratio for
         invariant masses 2--3~TeV. The acceptance interval is shorter.}
\label{fig:fig06}
\end{figure}

In Figure~\ref{fig:fig06} the rapidity distribution of the ratio
(DY+PI)/DY using MRST2004qed Mem1 for two invariant mass bins is
shown. The first bin is chosen to include the region where the ATLAS
and CMS collaborations have reported a hint for a diphoton excess in
the first Run~2 data. The PI effects are small for this bin, peaking
in the center of the detector, almost vanishing at rapidities
$|y|\sim$~2, and rising again at the acceptance edges. The rapidity
(y) distribution is symmetric around zero, as expected. This is a
good technical check that the calculation has been performed to high
enough precision, here to relative precision 0.1\%. If the
integration accuracy of the semi-analytic calculation is not high
enough, fluctuations and asymmetries begin to appear. The second bin
is a typical search bin from 2--3~TeV; here again the PI effects are
strongest in the center and are diminishing at the acceptance
edges. As the acceptance interval is shorter, the PI effects have no
room to rise again. The fact that at high mass the PI effects are
strongest in the center of the detector makes their suppression very
difficult. In addition, as these effects are small, especially if
MRST2004qed Mem1 is used, there appears to be no pressing need for
additional cuts, which could reduce the signal efficiency.

\begin{figure}[ht]
\centerline{\resizebox{0.95\textwidth}{10.5cm}{\includegraphics{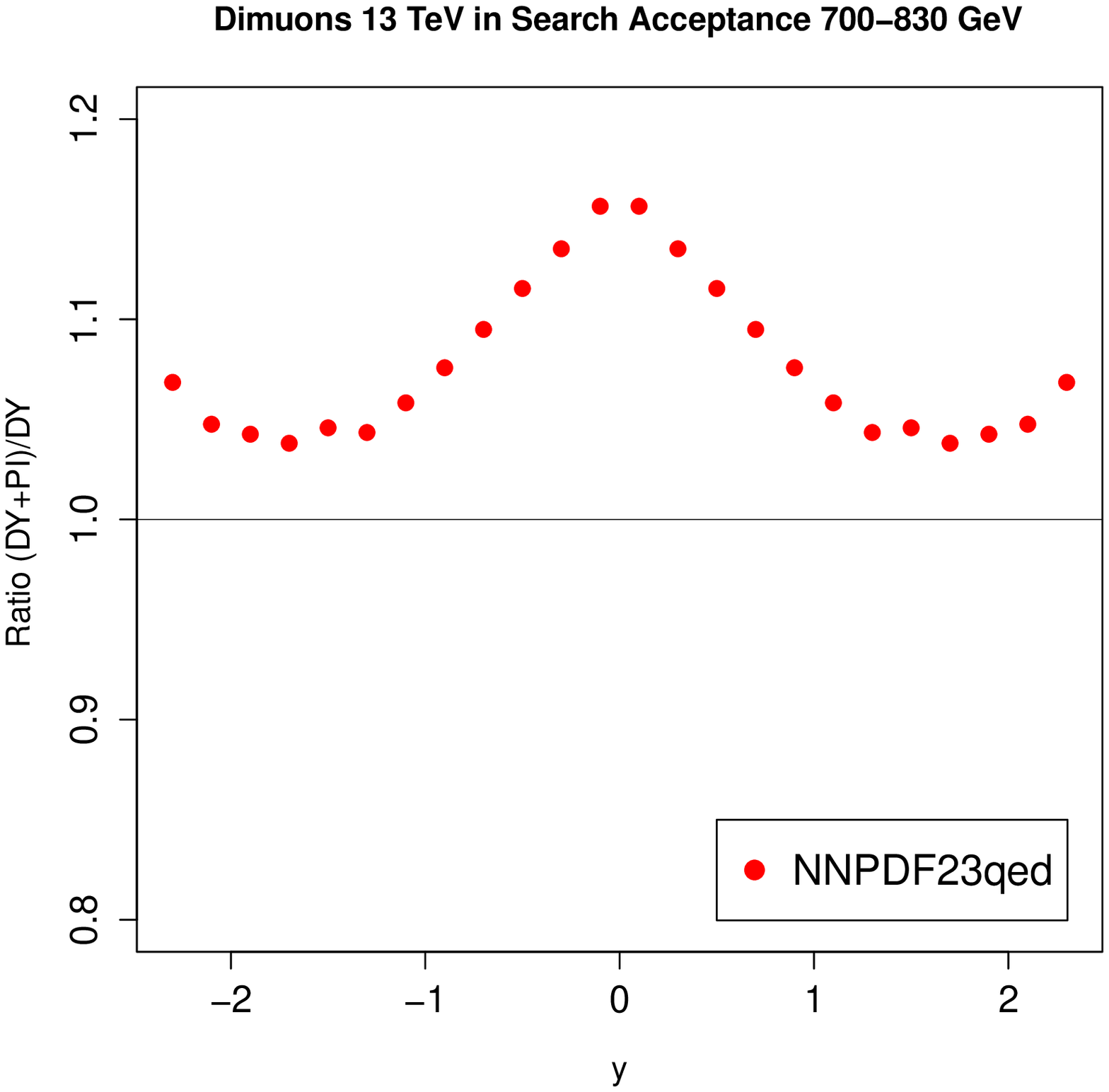}}}
\centerline{\resizebox{0.95\textwidth}{10.5cm}{\includegraphics{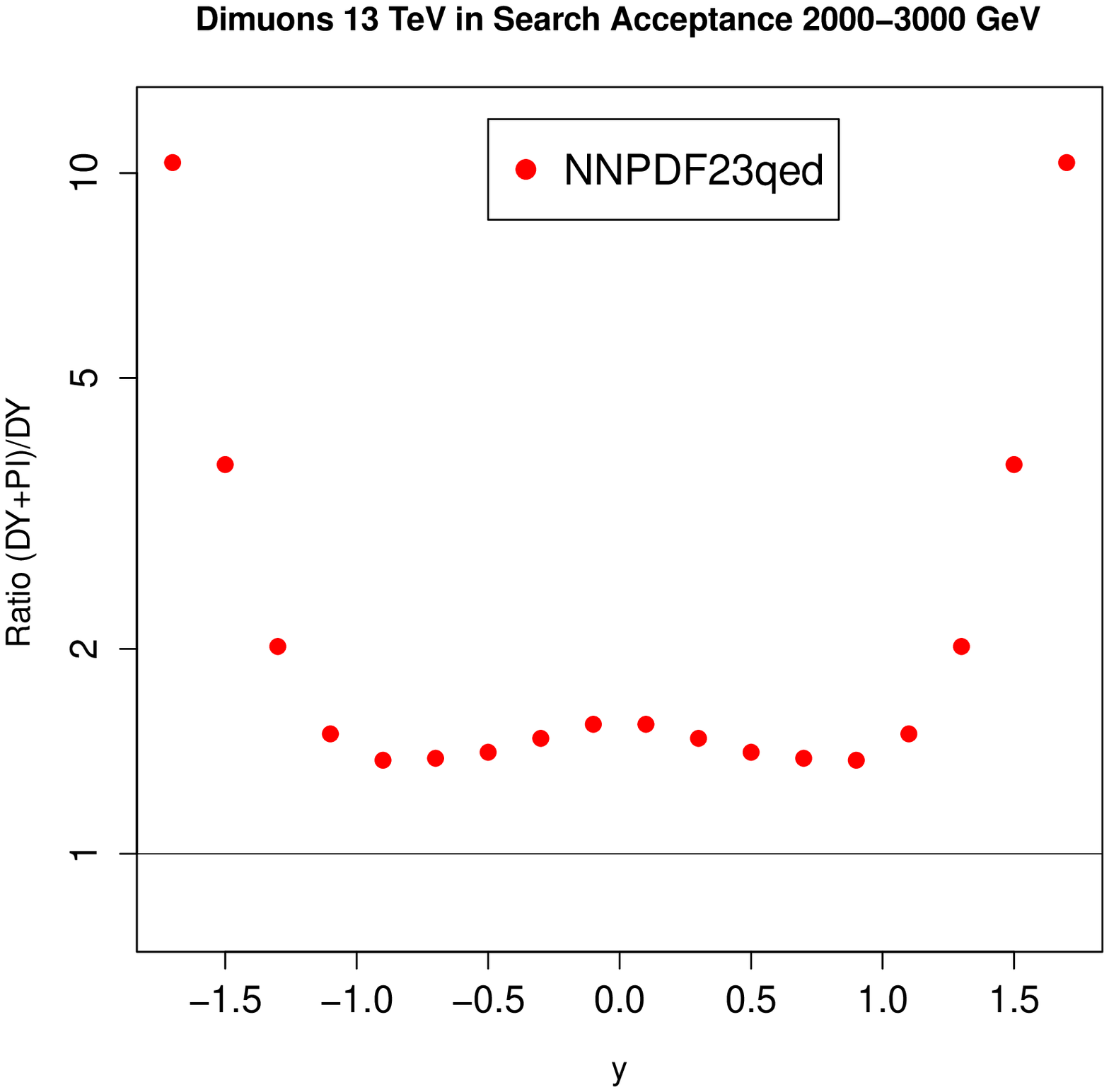}}}
\vspace*{-3pt}
\caption{Top: rapidity distribution for (DY+PI)/DY using NNPDF23qed
         for invariant masses 700--830~GeV.
         Bottom: rapidity distribution for the same ratio for
         invariant masses 2--3~TeV. Compare with Figure~\ref{fig:fig06}.}
\label{fig:fig07}
\end{figure}

In Figure~\ref{fig:fig07} the rapidity distribution of the ratio
(DY+PI)/DY using NNPDF23qed for the same two invariant mass bins is
shown. This time the PI effects are substantially higher for the first
bin, again peaking in the center of the detector, reduced at
rapidities $|y|\sim$~2, and rising again at the acceptance edges. For
the second bin the picture changes both qualitatively and
quantitatively.  A dramatic rise at the acceptance edges is seen, the
vertical scale for the ratio had to be increased from 1.2 to 10. The
lack of experimental and theory constraints result in the NNPDF23qed
becoming ``unhinged'' at high invariant masses. It is used in the
diphoton analysis of~\cite{Csaki:2016raa}. As noted
in~\cite{Ababekri:2016kkj}, the use of MRST2004qed Mem1 appears to
produce more realistic results.

The calculations of photon-induced effects in the dilepton channels
lead to the same conclusion: MRST2004qed or newer well behaved photon
PDFs are recommended for the exploration of the LHC search region at
high mass. With the accumulation of LHC luminosity the PI effects
around 750~GeV and higher can be measured in the dilepton channels,
improving the knowledge of the photon PDF.

\section{KISS}

\begin{figure}[ht]
\centerline{\resizebox{0.95\textwidth}{10.5cm}{\includegraphics{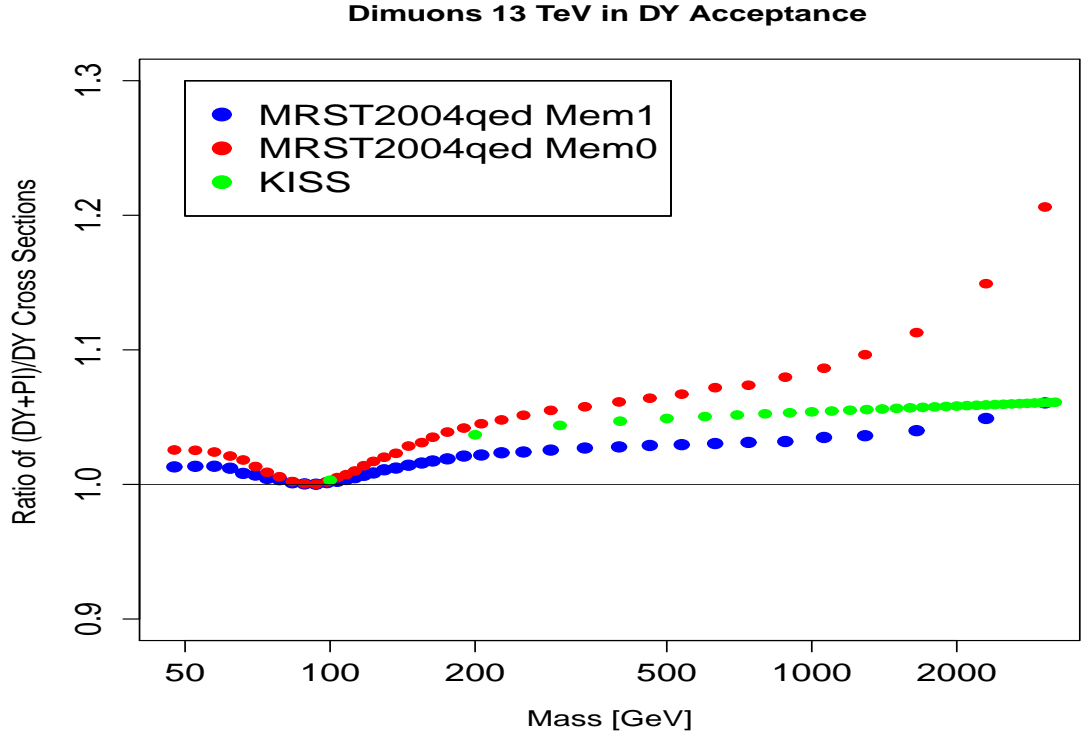}}}
\vspace*{-3pt}
\caption{Ratios of cross sections (DY+PI)/DY for MRST2004qed and the 
         KISS estimate in the DY acceptance. A simple approach
         produces estimates well within the MRST2004qed band.}
\label{fig:fig08}
\end{figure}

Before leaving this topic, it is tempting to compare the sophisticated
calculations presented above to a simple estimate. Assuming that the
cross section
ratio PI/DY is proportional to the QED and QCD coupling strengths
O($\alpha / \alpha_s)$, if only the photon exchange is considered for
Drell-Yan, and taking the renormalization group running of the
couplings to 2-loop into account, a rough estimate can be derived. In
order to obtain the full ratio, the Z exchange and $\gamma$Z
interference terms are added. The results are compared to the FEWZ
calculations with MRST2004qed in Figure~\ref{fig:fig08}. As is
evident, good agreement with the MRST2004qed predictions is observed,
with the KISS estimate well inside the band formed by the two PDF
members.

\section{Outlook}

The production of high invariant mass opposite sign lepton pairs in
proton-proton collisions at the LHC is an important search region for
manifestations of new physics, and for tests of the Standard Model at
highest momentum transfers. Gamma-gamma collisions, where photons
radiated by the incoming protons collide, will produce additional
lepton pairs - source of background for high mass resonant (like Z')
or non-resonant (like contact interactions) searches. In this paper
detailed calculations of the Drell-Yan and photon-induced cross
sections in the typical acceptance of a multi-purpose LHC detector at
center of mass energy 13~TeV are presented. The use of MRST2004qed in
the constituent mass option or newer well behaved photon PDFs is
recommended for the exploration of the LHC search region at high mass.
In addition, a simple estimate of the PI effects is shown to work
well from the Z peak to the high mass region.

The hint for a diphoton excess, reported by ATLAS and CMS at a mass
around 750~GeV, makes good theoretical understanding and measurements
in the dilepton channels at these energies very desirable in order to
explore the hypothesis that such an excess could be produced through
gamma-gamma collisions.

\vskip .5cm
\noindent
{\large{\bf{Acknowledgments}}}
\noindent

The author thanks the members of the University of Florida High Energy
Physics group for many productive discussions.

\clearpage

\end{document}